\pdfoutput=1
\documentclass[nologo,11pt,a4paper]{ETHpaper}

\usepackage{MnSymbol}
\usepackage[english]{babel}
\usepackage[numbers,sort&compress]{natbib} 
\usepackage{caption}

\usepackage{appendix}
\usepackage{xparse,tikz,calc}

\newcommand{\mean}[1]{\left\langle #1 \right\rangle}

\newcommand{\avg}[1]{\mean{#1}} \newcommand{\D}[2]{\frac{d #1}{d #2}} 
\usetikzlibrary{calc}
\tikzstyle{nosep}=[inner sep=0pt, outer sep=0pt]
\newcommand{\hcancel}[1]{\tikz[baseline=(tocancel.base)]{
        \node[nosep] (tocancel) {$#1$};
        \node[nosep, yshift=.5ex]  (from) at (tocancel.south west) {};
        \node[nosep, yshift=-.5ex] (to)   at (tocancel.north east) {};
        \draw[black] (from) -- (to);
    }}

\title{The ambiguous role of social influence on the wisdom of crowds: \\ An analytic approach}
\titlealternative{The ambiguous role of social influence on the wisdom of crowds: \\ An analytic approach}
\author{Pavlin Mavrodiev, Frank Schweitzer}
\authoralternative{P. Mavrodiev, F. Schweitzer}
\address{Chair of Systems Design, ETH Zurich, Switzerland\\
  {www.sg.ethz.ch}}
\www{{http://www.sg.ethz.ch}}

\reference{\emph{Submitted for publication}} 
\makeframing

\begin{document}

\maketitle

\begin{abstract}
  ``Wisdom of crowds'' refers to the phenomenon that the average opinion of a group of
  individuals on a given question can be very close to the true answer.
  It requires a large group diversity of opinions, but the collective error, the difference between the average opinion and the true value, has to be small.
  We consider a stochastic opinion dynamics where individuals can change their opinion based on the opinions of others (social influence $\alpha$), but to some degree also stick to their initial opinion (individual conviction $\beta$).
  We then derive analytic expressions for the dynamics of the collective error and the group diversity.
  We analyze their long-term behavior to determine the impact of the two parameters $(\alpha,\beta)$ and the initial opinion distribution on the wisdom of crowds.
  This allows us to quantify the ambiguous role of social influence: only if the initial collective error is large, it helps to improve the wisdom of crowds, but in most cases it deteriorates the outcome.
  In these cases, individual conviction still improves the wisdom of crowds because it mitigates the impact of   social influence. \\
  \emph{Keywords:} opinion dynamics, opinion distribution, collective effects, group decisions

\end{abstract}

\section{Introduction}

The idea to establish social science in the spirit of mathematics and physics dates back to the first half of the 19th century, when Auguste Compte (1798-1854) launched sociology based on the belief that the society follows general laws very much like the physical world.
At about the same time Adolphe Quetelet (1796-1874) published his "Essays on Social Physics" (1835), where he  applied probability theory to data from humans.
Later developments in sociophysics \citep{schweitzer2018} tried to adhere to these two approaches: derive a general dynamics applicable to societies, and analyze social data to find universal laws.
Our paper aligns to these developments. 

We do not enter the controversial discussion to what extent sociophysics has really contributed to the understanding of social systems. 
But a few conceptual frameworks from physics have indeed inspired the discussion about how to formalize social dynamics.
At the heart of statistical physics, as proposed in the late 19th century based on the fundamental works by Ludwig Boltzmann (1844-1906) and J. Williard Gibbs (1839-1903), is the problem of how the microscopic dynamics of system elements is linked to the dynamics of macroscopic system variables.
This question is of paramount importance also for the description of social and of economic systems.
For instance, how do the opinions of individuals contribute to the public opinion? How do decisions by individual consumers influence the market dynamics?

To answer such questions requires to solve a number of problems.
Ideally one wishes to have a theory that derives, from the proposed or observed dynamics of system elements (often termed as \emph{agents}), the macroscopic dynamics.
This works impressively well for some physical systems, but depends on a number of restrictive assumptions for socio-economic systems.
Nevertheless, the formalization of agent-based models, done in the right manner, is able to provide the \emph{micro-macro link} also for such systems, as for example the framework of Brownian agents witnesses \citep{agentbook-03}.

The second problem regards the identification of appropriate macroscopic variables that capture the relevant 
system dynamics.
Different aggregated measures can be proposed for socio-economic systems.
But do they really give insight into the conditions that determine the system dynamics?
Do they well reflect dependencies between individuals on the micro level?
Can they be linked to experiments or to empirical data?

In our paper, we discuss these problems for a specific application scenario, commonly denoted as the \emph{wisdom of crowds} (WoC) \citep{Page_2008,Hertwig_2012,Surowiecki2005}. 
In its simplest form, it describes a purely statistical effect.
If we take a \emph{large number} of \emph{independent} individual opinions on a certain question, the average over these opinions is surprisingly close to the \emph{true answer} to that question.
In particular, this average is closer to the truth than most individual and even expert opinions.
This finding was already reported by Francis Galton (1822-1911), who asked visitors at a life-stock fair about the weight of a particular ox \citep{Galton1907}.
Ever since, ample evidence for the WoC effect was provided for very  different scenarios, such as guessing tasks
\cite{Mannes2009}, problem-solving experiments \cite{Lorge1958337,Yaniv05usingadvice},
online communities \cite{Kittur2008,GarciaMartinez2014}, or prediction markets \citep{Ray2006}. 

At the same time there is evidence that the collective opinion of a group of individuals can be remarkably wrong with respect to an objective truth.
One reason for this comes from the fact that individual opinions are hardly \emph{independent}.
Instead, there are, often subtle, social influences that bias individual opinions.
In fact, it was shown that complex mechanisms of social contagion change not only the speed of opinion change, but also its outcome \citep{Karsai2014,fs-physa-02,Correia2018}.
On the systemic level, this \emph{can} lead to improvements for the wisdom of crowds, e.g. in case of social networks \citep{Becker2017}.
But more often, social influence in decision processes results into 
situations where, over time, the collective opinion moves farther and farther away from an objective truth, while all individuals are more and more convinced that they collectively move towards the right solution.
The wrong outcomes of such collective opinion dynamics can lead to disastrous situations as reported for financial bubbles and stock market crashes \cite{Malkiel2007,Eickhoff_2016}, panic stampedes \cite{JOHANSSON_2008}, the evaluation of the probability of an accident in NASA's shuttles \cite{Fisher2009}, or the famous
Stanford prison experiment \cite{Haney_1996}. It seems that the wisdom of crowds and the \textit{madness of crowds} \cite{Mackay2010,James_2018,Mollick2016} are two faces of the same coin.
This fuels the long-standing discussion under which conditions the wisdom of experts would outperform the wisdom of crowds \citep{Goldstein2014,Budescu2015,Navajas2018}.

In this paper, we aim at deriving \emph{formal expressions} for the systemic measures that are able to capture the effect of social influence on the WoC effect.
These measures, as proposed in the literature \citep{Page_2008,Lorenz2011a}, are the \emph{collective} error, measuring how close the average opinion is to the true value, and the \emph{group diversity}, measuring the variance of individual opinions.
For the wisdom of crowds, it is required that the group diversity is \emph{large}, while the collective error is \emph{small}. 
But it is to expect that, in the presence of social influence, the group diversity reduces drastically.
This bears the risk that individual opinions converge to a common opinion that is far way from the true value.
But there may be conditions under which social influence may help to converge to an opinion closer to the truth.

With our investigations, we want to better understand the ambiguous role of social influence. 
Our approach is motivated by (i) published experimental findings \citep{Lorenz2011a,Rauhut2011c} and (ii) their analysis by means of an agent-based model \citep{Mavrodiev2012a,mavrodiev2013}.
The novel contribution of this paper is in the derivation of \emph{analytical expressions} for the macroscopic quantities that shall describe the WoC effect.

So far, agent-based simulations have been used to obtain insights into the evolution of these quantities.
But the challenge to analytically solve the problem has not been tackled yet, also because it requires a lot of effort and some patience, as we demonstrate in this paper. 
Simulation results are usually quicker and more handy, but provide less insights into how the macroscopic indicators for the WoC effect are composed, and what they depend on.
Eventually, one wishes to quantify the range of parameters of the opinion dynamics that may lead to an enhancement or a deterioration of the WoC effect.

\section{Quantifying the wisdom of crowds}
\label{model-description}

\subsection{Measures for the wisdom of crowds}
\label{sec:macr-meas-wisd}

\paragraph{Opinion distribution. \ }

Let us consider an experimental situation, where subjects had, for instance, to estimate the length of the
border of Switzerland, which is a non-zero, positive and possibly large value \citep{Lorenz2011a}.
Each of these subjects returns an individual estimate, $x_{i}$, which we call an \emph{opinion} in the following.
The values of $x_{i}$ are strictly positive and very broadly distributed.
As the mentioned experiments have shown, the expected distribution of opinions, $P(x)$, is right-skewed and can be proxied by a \emph{log-normal   distribution}.
This means that the logarithms of $x_{i}$ follow a normal distribution
$\ln x_{i} \sim \mathcal{N}\left(\mu_{\ln x},\sigma^{2}_{\ln x}\right)$ with mean
$\mu_{\ln x}$ and variance $\mathrm{Var}[\ln x]= \sigma^{2}_{\ln x}$.

Because the underlying distribution is very broad, the \emph{average opinion} is not well represented by the arithmetic mean
\begin{align}
  \label{eq:11}
  \mu_{x}= \mean{x}=\sum\nolimits_{i=1}^{N}x_{i}
\end{align}
because it is much larger than most opinion values.
Instead, the geometric mean is the appropriate aggregation measure for the average opinion:
\begin{align}
  \label{eq:8}
  \mu_{\mathrm{geo}}= \left\{\prod\nolimits_{i=1}^{N}x_{i}\right\}^{1/N} \;\Rightarrow \;
  \mu_{\ln x} = \ln \mu_{\mathrm{geo}} = \sum\nolimits_{i=1}^{N} \ln x_{i} = \mean{\ln x}
\end{align}

In the following, we consider that subjects can change their opinion over time either because of random influences or in response to information received from other subjects.
The details of these influences are specified in the following section.
But from now on, all quantities become time-dependent, i.e. opinions become $x_{i}(t)$.
While their initial values $x_{i}(0)$ follow a log-normal distribution, the same does not hold for times $t>0$, because of the assumed influences on the change of opinions.

The wisdom of crowds is expected to work if the diversity of individual opinions is
\emph{large}, while the deviation of the average opinion from the true value $\mathcal{T}$ is
\emph{small}.  Therefore, in line with previous studies \citep{Lorenz2011a}, we will use the
\textit{group diversity}, $\mathcal{D}(t)$, and the \textit{collective error}, $\mathcal{E}(t)$, as
macroscopic measures to evaluate these conditions.
We are particularly interested in the time-dependent change and the stationary values of these measures, as they can indicate under which conditions the wisdom of crowds will break down.
To analyze these conditions is the aim of our paper.

\paragraph{Collective error. \ }

$\mathcal{E}(t)$ shall be defined as the squared deviation of the average opinion from the true value,
$\mathcal{T}$:
\begin{equation}
  \label{CE}
  \mathcal{E}(t) = \left[\ln{\mathcal{T}} - \avg{\ln{x(t)}} \right]^{2},
\end{equation}
For the dynamics follows:
\begin{align}
  \label{dCE}
  \D{}{t}\mathcal{E}(t) &= -2\left[\ln \mathcal{T} - \avg{\ln
                          x(t)}\right]\, \D{}{t}\avg{\ln x(t)}
\end{align}

\paragraph{Group diversity. \ }

We express $\mathcal{D}(t)$ by the \emph{variance} of the opinion distribution:
\begin{align}
  \label{Diversity}
  \mathcal{D}(t) &= \mathrm{Var}\left[\ln x(t)\right]= \dfrac{1}{N}\displaystyle\sum\limits_{i=1}^{N}\left[ \ln
                   x_{i}(t) - \avg{\ln x(t)}\right]^{2} = \avg{\left[\ln x(t)\right]^{2}} - \avg{\ln
                   x(t)}^{2}
\end{align}
In order to derive a dynamic equation for the group diversity we use the delta method to approximate
the variance.  This method is in essence a first-order Taylor expansion of the form:
\begin{align}
  \mathrm{Var}\left[f(X)\right] \approx \left[f'\left(\avg{X}\right)\right]^{2}\mathrm{Var}[X]
  \label{eq:6}
\end{align}
The method will be a poor approximation in cases where $f(X)$ is highly non-linear.  This is not the
case when $f(X)=\ln X$. For the calculation we write
\begin{align}
  x_{i}(t)=\avg{x(t)}+\delta_{i}(t)
  \label{eq:9}
\end{align}
where $\delta_{i}(t)$ is an individual's deviation from the average opinion, with
$\avg{\delta(t)}=0$.  With this notation and Eqn.~\eqref{eq:6}, the group diversity becomes:
\begin{align}
  \label{eq:7}
  \mathcal{D}(t) = \dfrac{1}{\avg{x(t)}^{2}}\avg{\delta^{2}(t)}
\end{align}
This compact expression allows us to derive the dynamics in the form:
\begin{align}
  \label{Diversity3}
  \D{}{t}\mathcal{D}(t)&=\dfrac{1}{\avg{x(t)}^{2}}\D{}{t}\avg{\delta^{2}(t)}
                         -
                         \dfrac{2}{\avg{x(t)}^{4}}\avg{\delta^{2}(t)}\avg{x(t)}\D{}{t}\avg{x(t)}
\end{align}

\subsection{Opinion dynamics}
\label{sec:micro-dynam-opin}

To formalize how the opinion $x_{i}(t)$ of each subject changes over time, we build on the framework of \emph{Brownian agents} \citep{agentbook-03}, which considers a superposition of deterministic and stochastic influences on the dynamics: 
\begin{align}
  \label{eq:1}
  \frac{d x_{i}(t)}{dt}= - \beta \left[x_{i}(t) -x_{i}(0)\right] + \frac{1}{N} \sum_{j} \mathcal{F}_{ij}(t) + A\xi_{i}(t)
\end{align}
$x_{i}(0)$ denotes the initial value.
The parameter $\beta$ describes the \emph{individual conviction} about the own opinion.
The larger $\beta$, the more an agent tries to stick to the initial opinion.
$\xi_{i}(t)$ is Gaussian white noise, i.e. it is not correlated in time, $\mean{\xi_{i}(t)\xi_{i}(t^{\prime})}=\delta(t-t^{\prime})$, and zero on average, $\mean{\xi_{i}(t)=0}$. 
$A$ denotes the strength of the stochastic force. 
The term $\mathcal{F}_{ij}(t)$ eventually describes how the change of opinion of agent $i$ is
influenced by the opinion of other agents $j$.
Here, we assume that agents have information only about the \emph{average opinion} of all other
agents, which is equivalent to a mean-field scenario.  This is reflected in the following assumption
for $\mathcal{F}_{ij}(t)$:
\begin{align}
  \mathcal{F}_{ij}(t)= \alpha\, \left[x_{j}(t)-x_{i}(t)\right]
  \label{eq:4}
\end{align}
The parameter $\alpha$ describes the strength of the social influence of other opinions $x_{j}(t)$ on the opinion $x_{i}(t)$.
In our ansatz the social influence from other opinions increases with the difference
between opinions.
While this sounds like a simplified assumption, it has been empirically justified
in \citep{mavrodiev2013}, therefore we use it here.
Because the coupling variable $\alpha$ is effectively a
\emph{constant}, equal for all $i$, we have:
\begin{align}
  \frac{1}{N}  \sum_{j=1}^{N} \mathcal{F}_{ij}(t)=  \alpha \left[ \mean{x(t)}-x_{i}(t)\right]\;;\quad
  \mean{x(t)}=  \dfrac{1}{N}  \sum_{j=1}^{N} x_{j}(t)
  \label{eq:5}
\end{align}
where $\mean{x(t)}$ is denoted as the \emph{mean} opinion in the following.
Eqn.~\eqref{eq:1} therefore results in the stochastic dynamics:
\begin{equation}
  \label{estimates}
  \D{x_{i}(t)}{t} = \alpha \left [\avg{x(t)} - x_{i}(t) \right] + \beta  \left[x_{i}(0) - x_{i}(t) \right] + A\xi_{i}(t)
\end{equation}
Averaging the dynamics of
Eq.~\eqref{estimates} over the whole agent population yields a simple linear form:
\begin{equation}
  \label{estimates2}
  \D{\avg{x(t)}}{t} = \beta\left[\avg{x(0)} - \avg{x(t)}\right] + \dfrac{A}{\sqrt{N}}\avg{\xi(t)}
\end{equation}
which is a standard Ornstein-Uhlenbeck process with the solution:
\begin{equation}
  \label{estimates2solution}
  \avg{x(t)} = \avg{x(0)}e^{-\beta t} + \avg{x(0)}\left(1-e^{-\beta t}\right) + \dfrac{A}{\sqrt{N}} \displaystyle \int_{0}^{t}e^{\beta(s-t)}\avg{\xi(s)}ds
  \end{equation}
Therefore the \emph{time average} of
the \emph{ensemble average} of the opinons, $\overline{\avg{x(t)}}$, equals $\avg{x(0)}$ for large
$t$.

\paragraph{Initial configuration. \ }

In order to solve our dynamic equations for the macroscopic measures $\mathcal{D}(t)$ and $\mathcal{E}(t)$, we still need to
specify the initial distribution of opinions.  We take the log-normal distribution $P(x,0)$ with the
parameters $\mu_{\ln x}(0)$ and $\sigma^{2}_{\ln x}(0)$ as an input, from which $N$ values
$x_{i}(0)$ are sampled.  Each of these initial values can be represented as
$x_{i}(0)=\mean{x(0)}+\delta_{i}(0)$, Eqn.~\eqref{eq:9}, where $\delta_{i}(0)$ is the deviation of the initial opinion
from the \emph{initial mean} $\mean{x(0)}$.  By definition $\mean{\delta(0)}=0$.  We note that only
the  $\delta_{i}(0)$ result
 from the log-normal distribution, while in Eqn.~\eqref{eq:9} the $x_{i}(t)$ and $\delta_{i}(t)$ are determined by Eqn.~\eqref{estimates}.
 
The initial collective error $\mathcal{E}(0)$ and the initial average opinion are related by
Eqn.~\eqref{CE}:
\begin{align}
  \label{logestimateszero}
  \mu_{\ln x}(0) = \avg{\ln x(0)} = \ln \mathcal{T} \pm \sqrt{\mathcal{E}(0)}
\end{align}
Hence, we only need an additional value $\sigma^{2}_{\ln x}(0)$, to calculate the initial group
diversity $\mathcal{D}(0)$.  Then, the initial configuration on the \emph{macroscopic} level is
given by the pair $\{\mathcal{E}(0)$,$\mathcal{D}(0)\}$.  It will be of interest to us to study the
dynamics of these values, in particular their long-term values for $t\to \infty$,
$\{\mathcal{E}_{\mathrm{LT}},\mathcal{D}_{\mathrm{LT}}\}$.

With these specifications of the systemic measures and the agent variables, we now proceed solving the dynamics  analytically.

\section{Analytical results}
\label{sec:analytical-results}

\subsection{Collective error}

\paragraph{Dynamic solution. \ }

To calculate $\mathcal{E}(t)$, Eqn.~\eqref{CE} and $d \mathcal{E}(t)/d t$, Eqn.~\eqref{dCE}, we need to have the explicit expressions for the following quantities: $\mean{\ln x(t)}$, $d \mean{\ln x(t)}/dt$.
These are derived in Appendices~\ref{dlogestimates2appendix} and \ref{sec:derivation-eqn}. 
Here we only present the results:
\begin{align}
  \label{eq:dlnestimates}
  \D{\avg{\ln x(t)}}{t} = &
                          (\alpha+\beta)\left[1-\dfrac{\beta}{\beta+\alpha e^{-(\alpha+\beta)t}}\right]\  \sum_{n=1}^{\infty}
                          \dfrac{(-1)^{n}}{\avg{x(0)}^{n}} \dfrac{\avg{\delta^{n}(0)}}{(\alpha+\beta)^{n}}\big[\beta+\alpha
                          e^{-(\alpha+\beta)t}\big]^{n} \nonumber \\
                        &  +\dfrac{A}{\sqrt{N}}\dfrac{\avg{\xi(t)}}{\avg{x(0)}}
\end{align}
Integration leads to the  solution:
\begin{align}
  \label{eq:3}
  \avg{\ln x(t)} &= \avg{\ln x(0)} + \sum_{n=1}^{\infty}
                   \dfrac{(-1)^{n}}{\avg{x(0)}^{n}}
                   \dfrac{\alpha^{n}\avg{\delta^{n}(0)}}{n(\alpha+\beta)^{n}}\
                   \left[1-e^{-(\alpha+\beta)nt}\right]+
  \\ \nonumber
                 &\quad + \sum_{n=1}^{\infty}
                   \dfrac{(-1)^{n}}{\avg{x(0)}^{n}}
                   \dfrac{\avg{\delta^{n}(0)}}{(\alpha+\beta)^{n}}\
                   \sum_{k=1}^{n-1}\dfrac{1}{k}{n-1 \choose
                   k-1} \ \beta^{n-k}\alpha^{k}
                   \left[1-e^{-(\alpha+\beta)kt}\right]
\end{align}
With these expressions we have completely described the dynamics of the collective error.

\paragraph{Asymptotic solution. \ }

The long term behavior of the collective error results from $t\to \infty$, and we find
\begin{align}
  \label{eq:36}
  \mathcal{E}_{\mathrm{LT}}&= \left[\ln \mathcal{T} - \avg{\ln x(0)} - \sum_{n=1}^{\infty}
                   \dfrac{(-1)^{n}}{\avg{x(0)}^{n}}
                   \dfrac{\alpha^{n}\avg{\delta^{n}(0)}}{n(\alpha+\beta)^{n}}\
                  \right.  \\ \nonumber
                 &\quad \left. - \sum_{n=1}^{\infty}
                   \dfrac{(-1)^{n}}{\avg{x(0)}^{n}}
                   \dfrac{\avg{\delta^{n}(0)}}{(\alpha+\beta)^{n}}\
                   \sum_{k=1}^{n-1}\dfrac{1}{k}{n-1 \choose
                   k-1} \ \beta^{n-k}\alpha^{k}
                                     \right]^{2}
\end{align}
As we see, the \emph{final} outcome of the collective error is mainly determined by the properties of the \emph{initial opinion distribution}, in particular $\mean{x(0)}$ and $\mean{\delta(0)}$, and further by the two model parameters social influence $\alpha$ and individual conviction $\beta$.
We will discuss the consequences of this in the next Section.

\subsection{Group diversity}
\label{sec:group-diversity}

\paragraph{Dynamic solution. \ }

To calculate $\mathcal{D}(t)$, Eqn.~\eqref{eq:7}, and $d\mathcal{D}(t)/dt$, Eqn.~\eqref{Diversity3}, we need to have the explicit expressions for $\mean{x(t)}$ and $\mean{\delta^{2}(t)}$ as well as their time derivatives.  
Again, here we only present the result, the derivation is provided in Appendix~\ref{deltasqappendix}.
Plugging in $\avg{\delta^{2}(t)}$ from Eq. \ref{eq:dtn} into $\mathcal{D}(t)$, Eqn.~\eqref{eq:7},  yields:
\begin{align}
  \label{eq:diversity}
  \mathcal{D}(t) =
  \dfrac{\avg{\delta^{2}(0)}}{[\avg{x(t)}(\alpha+\beta)]^{2}}[\beta+\alpha
  e^{-(\alpha+\beta)t}]^{2}  
\end{align}
which is always positive.

\paragraph{Asymptotic solution. \ }

The long term behavior of the group diversity results from $t\to \infty$, and we find
\begin{align}
  \label{eq:diversity-long-term}
  \mathcal{D}_{\text{LT}} =
  \dfrac{\avg{\delta^{2}(0)}\beta^{2}}{[\avg{x(0)}(\alpha+\beta)]^{2}}
\end{align}
where we made use of the fact that $\mean{x(t)}\approx \mean{x(0)}$, see Eqn.~\eqref{estimates2solution}.

We can further derive how the long-term group diversity depends on the two model parameters, social influence $\alpha$ and individual conviction $\beta$:
\begin{equation}
  \label{diversity-alpha}
  \D{}{\alpha}\mathcal{D}_{\text{LT}} = -\dfrac{2\avg{\delta^{2}(0)}\beta^{2}}{\avg{x(0)}^{2}(\alpha+\beta)^{3}}  < 0 
\;; \quad 
  \D{}{\beta}\mathcal{D}_{\text{LT}} = \dfrac{2\alpha\beta\avg{\delta^{2}(0)}}{\avg{x(0)}^{2}(\alpha+\beta)^{3}}  > 0     
\end{equation}
This will be tested in the next section.
Further, we see that the \emph{final} group diversity increases with the \emph{initial} deviation from the mean,  $\avg{\delta^{2}(0)}$:
\begin{equation}
  \label{diversity-variance}
  \D{}{\avg{\delta^{2}(0)}}\mathcal{D}_{\text{LT}} = \dfrac{\beta^{2}}{\avg{x(0)}^{2}(\alpha+\beta)^{2}}  > 0     
\end{equation}

\section{Discussion}
\label{sec:discussion-1}

\subsection{Collective error}
\label{sec:collective-error}

We now use our analytical solutions from the previous section to study the dependence of the collective error and the group diversity on the two model parameters, social influence $\alpha$ and individual conviction $\beta$, and on the variance of the initial opinion distribution.

\begin{figure}[htbp]
  \includegraphics[scale=0.33]{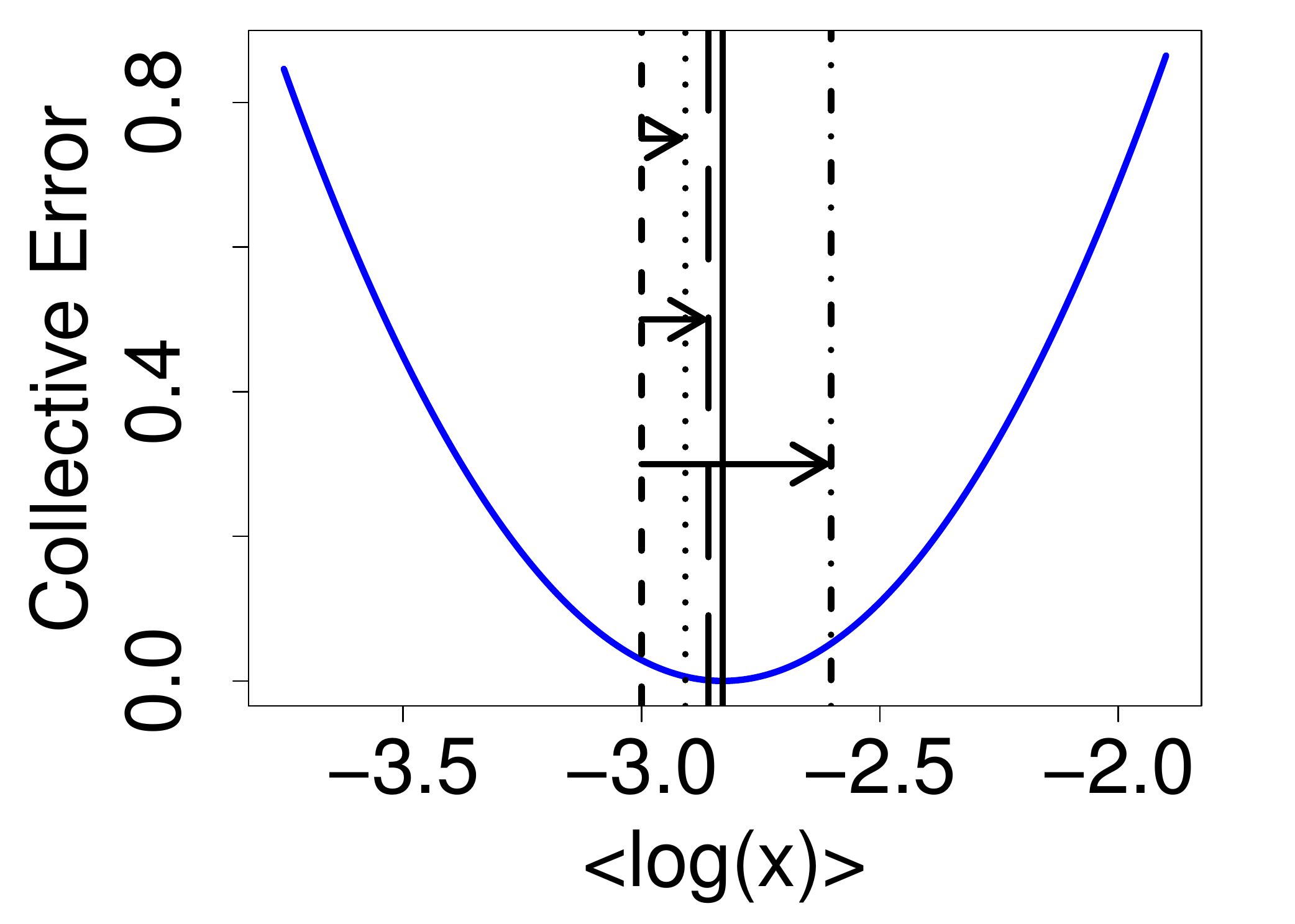}(a) \hfill
  \includegraphics[scale=0.33]{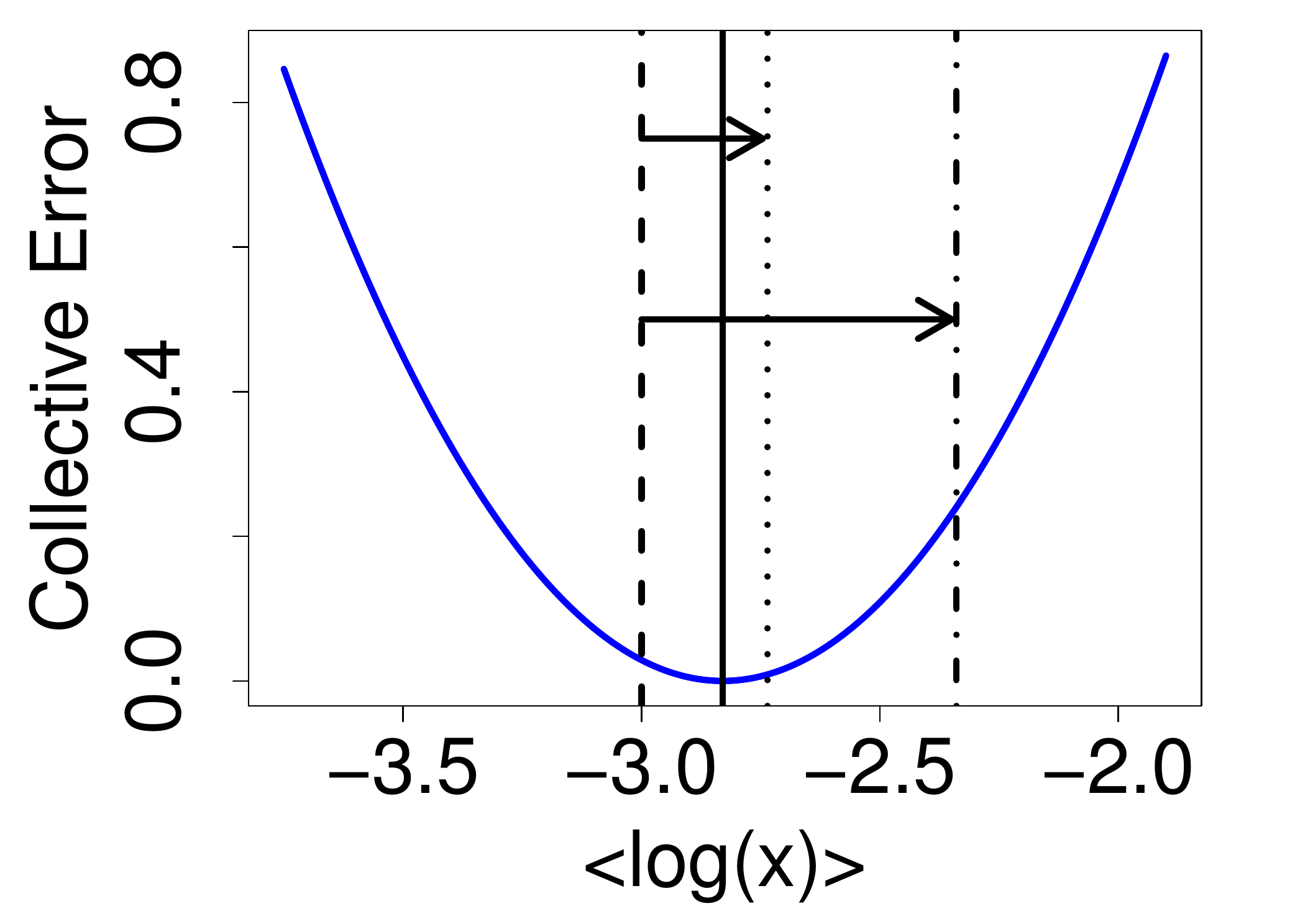}(b)
  \caption{ Collective error $\mathcal{E}$ dependent on $\mean{\ln x}$.  True value:
    $\ln \mathcal{T}=-2.83$ (solid black line).  Initial condition: $\avg{\ln x(0)}=-3$ (dashed
    black line), $\mathcal{E}(0)=0.02$, $\avg{x(0)}=0.075$. \\
    \textbf{(a)} $\avg{\delta^{2}(0)}=0.006$.  Stationary states for \emph{different parameters}
    $(\alpha,\beta)$: (0.01, 0.1) (dotted line),
    (0.1, 0.1) (dotted-dashed line), (0.1, 0.6) (long-dashed line). \\
    \textbf{(b)} $(\alpha,\beta)=(0.1, 0.1)$.  Stationary states for \emph{different initial
      variances} $\avg{\delta^{2}(0)}$: 0.004 (dashed line), 0.01 (dotted-dashed line).  }
  \label{cestationary23}
\end{figure}

The collective error $\mathcal{E}(t)$ is defined as the \emph{squared} difference between the true value, $\ln \mathcal{T}$, and the mean opinion, $\mean{\ln x(t)}$.
Hence, the plot gives a parabola with the minimum at $\ln \mathcal{T}=\mean{\ln x(t)}$.
The asymptotic value $\mean{\ln x_{\mathrm{LT}}}$ is calculated from Eqn.~\eqref{eq:3}. 
Figure~\ref{cestationary23} illustrates how $\mean{\ln x_{\mathrm{LT}}}$ and consequently $\mathcal{E}_{\mathrm{LT}}$, Eqn.~\eqref{eq:36}, depend on the two parameters $(\alpha,\beta)$ (a) and on the initial variance $\avg{\delta^{2}(0)}$ (b).

Comparing the sets of parameters $(0.01,0.1)$ and $(0.1,0.1)$ we see that for smaller \emph{social influence} $\alpha$  the motion of $\avg{\ln x(t)}$ is slower  and the quasi-stationary state of $\mean{\ln x}$ is respectively closer to the initial value (Figure \ref{cestationary23}(a), dotted vs. dotted-dashed line).
On the other hand, comparing the sets of parameters $(0.1,0.1)$ and $(0.1,0.6)$ we see that for smaller \emph{individual conviction} $\beta$  the motion of $\avg{\ln x(t)}$ is larger  and the quasi-stationary state of $\mean{\ln x}$ is respectively further away from the initial value (Figure \ref{cestationary23}(a), dotted-dashed vs. long-dashed line).

Obviously, a larger individual conviction $\beta$ counteracts the effect of a larger social influence $\alpha$, which ultimately results in the smallest collective error.
That means, we observe a struggle between social influence $\alpha$ and individual conviction $\beta$.
A larger $\alpha$ expands the range of motion of $\avg{\ln x(t)}$, consequently increases the convergence
limit, whereas a larger $\beta$ restricts it.
These opposing effects \emph{may} lead to a reduction of long-term $\mathcal{E}$, as in Figure
\ref{cestationary23}(a).
But, as we will demonstrate in Figure~\ref{sweep} in the next Section, 
conditional on the initial conditions $\mathcal{E}(0)$ and $\mathcal{D}(0)$, these two opposing effects \emph{may} also lead to a \emph{deterioration}. 

The effect of the initial variance, $\avg{\delta^{2}(0)}$, on the long-term collective error is similar to the influence of $\alpha$.
A larger heterogeneity in the initial opinions which directly translates to larger initial group diversity, leads to
longer motion of $\avg{\ln x(t)}$ (Figure \ref{cestationary23}(b), dotted vs. dotted-dashed
lines) and  consequently to a higher
$\avg{\ln x_{\mathrm{LT}}}$.
In line with the above argumentation, whether this has a positive or a negative
net effect on the collective error depends entirely on the initial distribution of opinions and is further discussed in Section~\ref{sec:sweep}.

In conclusion, for given parameters $(\alpha,\beta)$ the initial condition $\avg{\ln x(0)}$ uniquely determines the end value of the collective error.
It should be noted that the
log of the geometric mean exhibits only rightward motion, since $d{\avg{\ln x(t)}}/{dt}>0$, Eqn.~\eqref{eq:dlnestimates}.
Hence, if we start from $\avg{\ln x(0)} > \ln (\mathcal{T})$ the collective
error will always increase.

\subsection{Group diversity}
\label{sec:group-diversity-1}

\begin{figure}[htbp]
\includegraphics[scale=0.33]{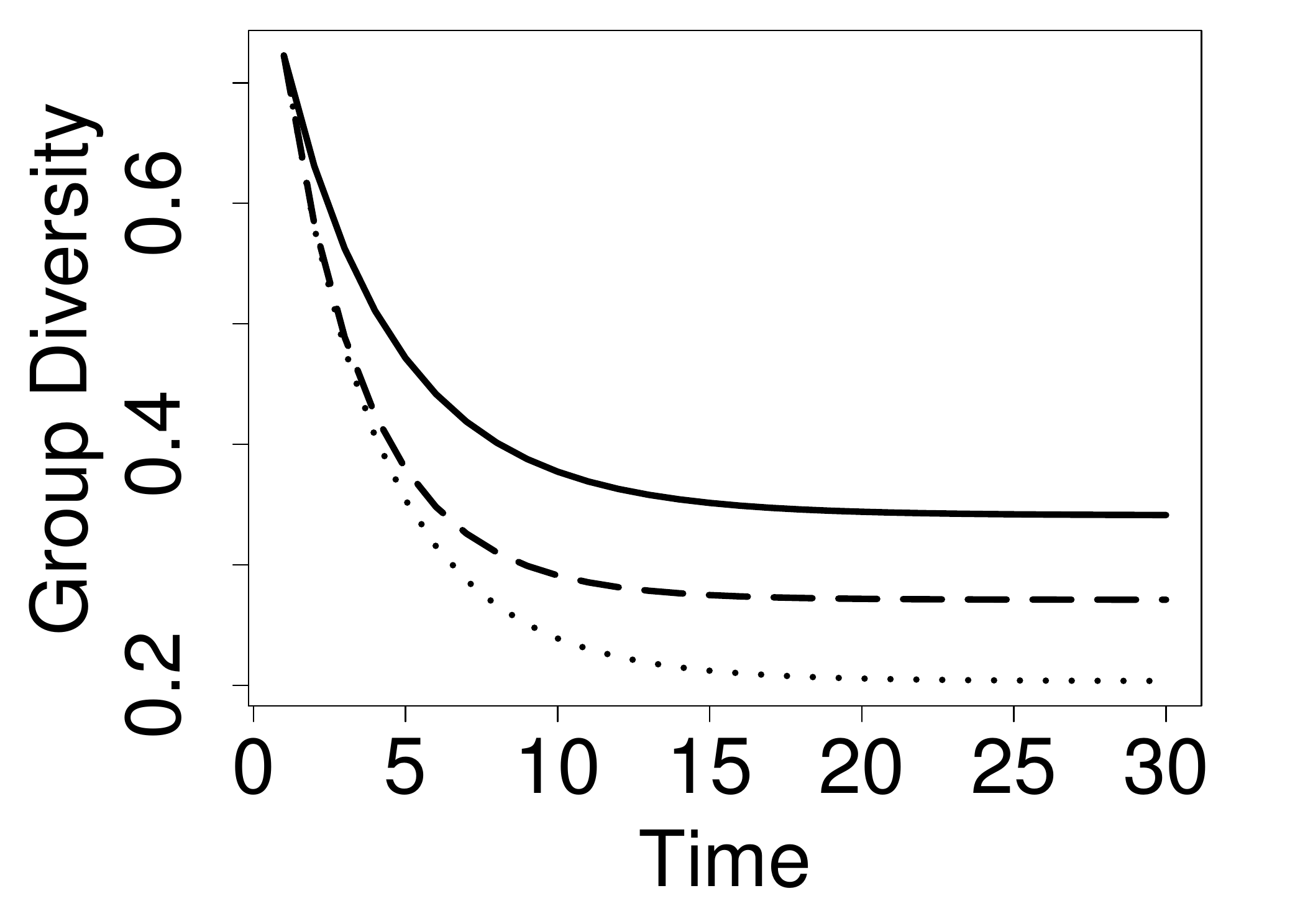}(a)
\hfill
\includegraphics[scale=0.33]{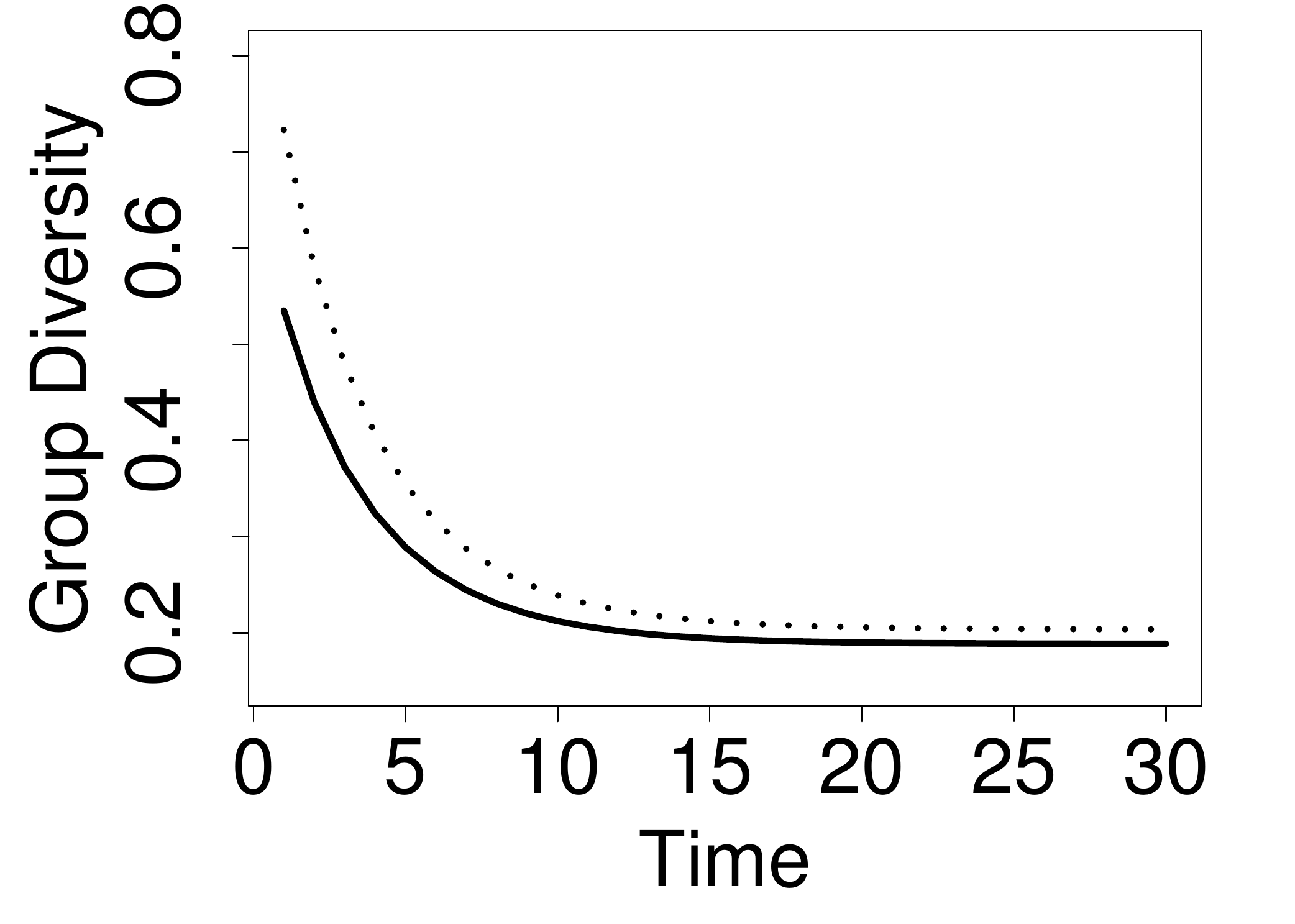}(b)
  \caption{Evolution of the group diversity $\mathcal{D}(t)$ over time. \\
    \textbf{(a)}  $\avg{\delta^{2}(0)}=0.006$, $\mathcal{D}(0)=0.8$, $\avg{x(0)}=0.075$.
    Different \emph{parameters} $(\alpha,\beta)$:
    (0.5, 2.0) (solid line), (0.8, 2.0) (dotted line), (0.8, 2.5) (dashed line). \\
    \textbf{(b)} $(\alpha,\beta)=(0.8,0.2)$, $\avg{x(0)}=0.075$. Different \emph{initial variances} $\avg{\delta^{2}(0)}$:
    0.006 (dotted line), 0.004 (solid line).
} 
  \label{diversitystationary23}
\end{figure}

The dependency of the group diversity $\mathcal{D}(t)$ on the parameters $(\alpha,\beta)$ and the initial variance $\avg{\delta^{2}(0)}$ resembles that of the collective error (within the parameter space in which it is well defined).
Figure \ref{diversitystationary23}(a) shows the effect of the social influence $\alpha$ and the individual conviction $\beta$.
As expected from the analytical solution in Eqn.~\eqref{diversity-alpha}, the increase of social influence leads to a gradual decrease of the group diversity towards the steady state given by Eqn. \eqref{eq:diversity-long-term}, whereas the increase of individual conviction leads to an increase in group diversity.

Figure \ref{diversitystationary23}(b) illustrates that an increase of the initial variance slightly increases also the group diversity.
However, because of the social influences, the impact of the initial variance becomes smaller over time.

\subsection{Exploring the parameter space}
\label{sec:sweep}

\begin{figure}[htbp]
  \includegraphics[width=0.29\textwidth]{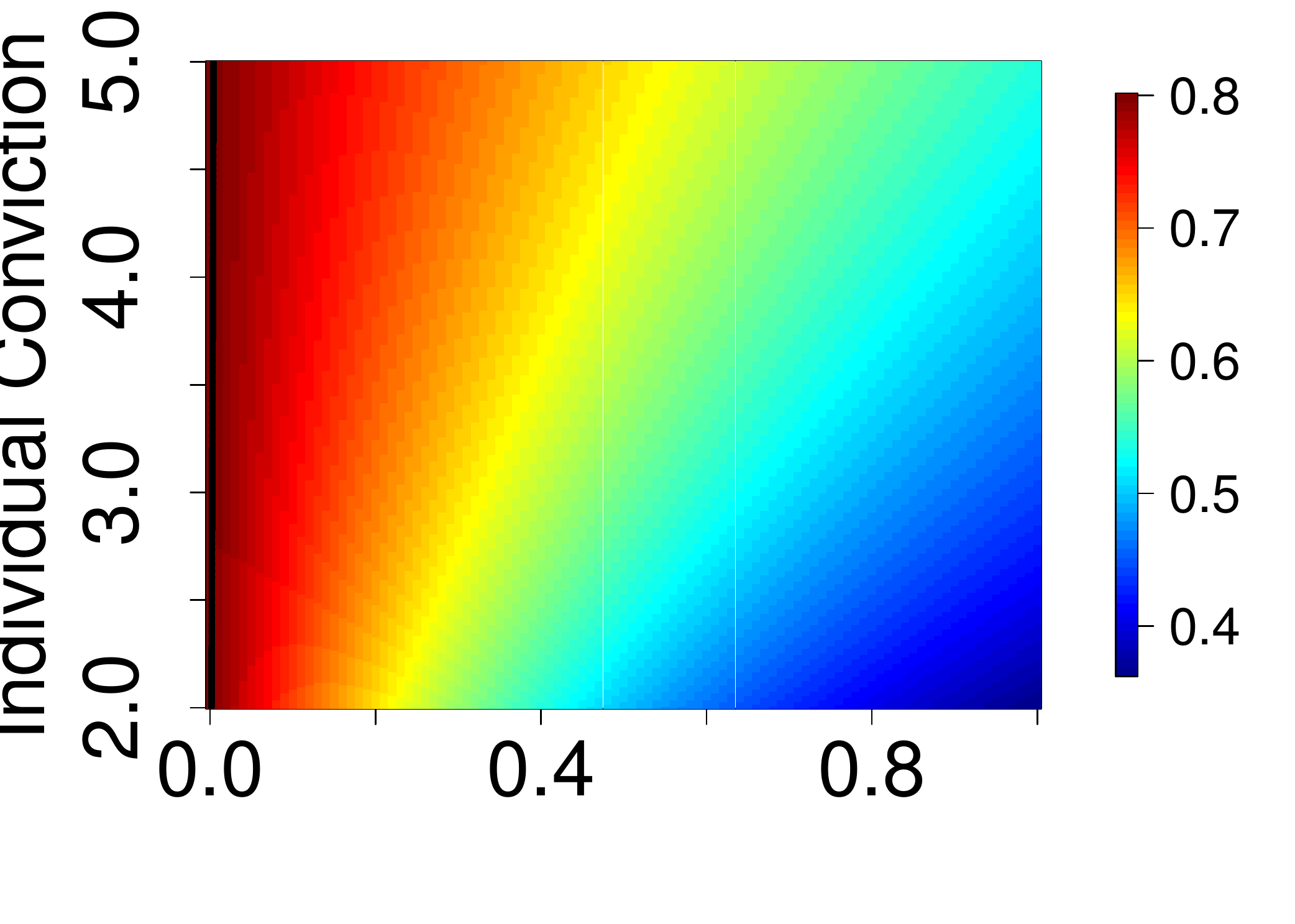}(a)
    \hfill
    \includegraphics[width=0.29\textwidth]{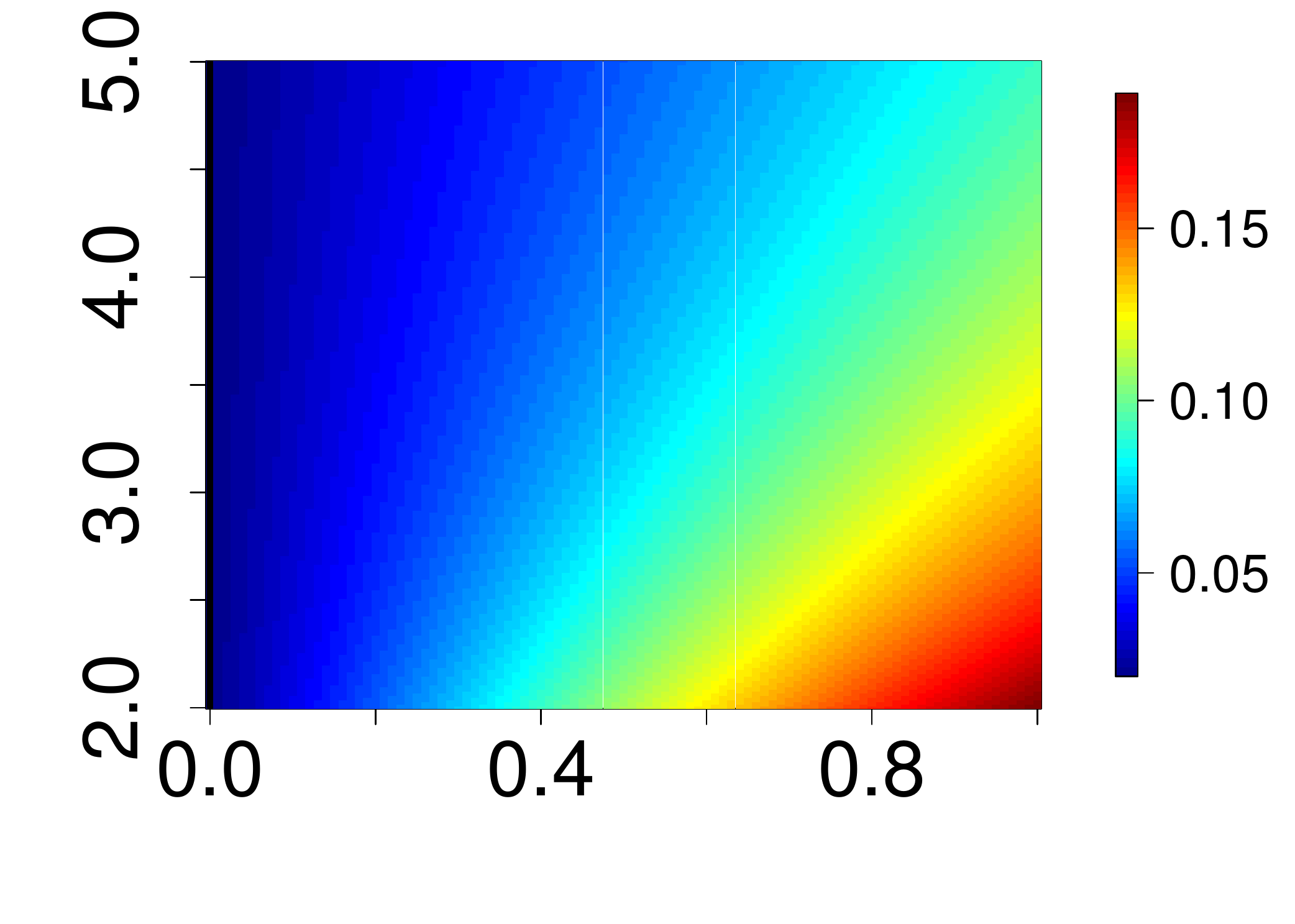}(b)
      \hfill
\includegraphics[width=0.29\textwidth]{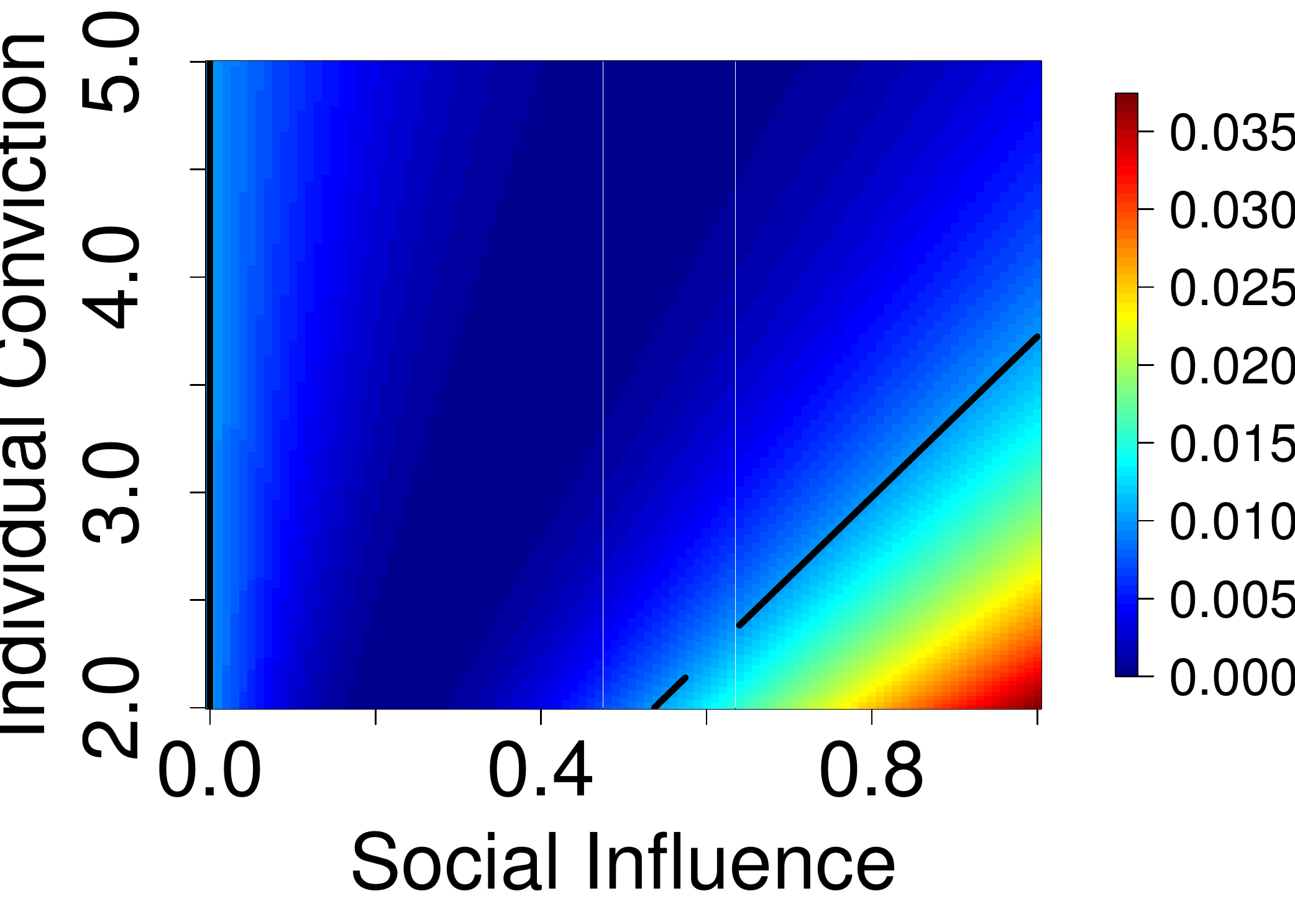}(c)
\caption{Color-coded long-term collective error $\mathcal{E}_{\mathrm{LT}}$ dependent on the values of social influence $\alpha$ ($x$-axis) and individual conviction $\beta$ ($y$-axis). 
  Different initial conditions:\\
  \textbf{(a)} $\mathcal{E}$(0)=0.80, $\ln \mathcal{T}$=-2.00, $\avg{\ln x(0)}$=-2.9, \\
  \textbf{(b)} $\mathcal{E}$(0)=0.02, $\ln \mathcal{T}$=-3.12, $\avg{\ln x(0)}$=-3.0, \\
  \textbf{(c)} $\mathcal{E}$(0)=0.01, $\ln \mathcal{T}$=-2.90, $\avg{\ln x(0)}$=-3.0.\\
  Black contour lines indicate regions in the parameter space where $\mathcal{E}_{\mathrm{LT}}=\mathcal{E}(0)$.
  Note that these are vertical lines at $\alpha$=0 in all plots, in (c) there is an additional line. 
}
  \label{sweep}
\end{figure}

To complement the discussion of the specific cases above and to reveal a more profound insight on the impact of social influence  and individual conviction, we run a parameter sweep on $(\alpha,\beta)$. 
In Figure \ref{sweep} we have calculated the long-term collective error $\mathcal{E}_{\mathrm{LT}}$, Eqn.~\eqref{eq:36}, for three different initial values of the collective error which are chosen such that they represent three characteristic scenarios. 
The color code indicates the value of $\mathcal{E}_{\mathrm{LT}}$, (blue) for \emph{low} values, which is positive, (red) for \emph{high} values, which is negative.
We note that for each plot, the color code represents \emph{different} values, and (blue) not always means $\mathcal{E}_{\mathrm{LT}}=0$. 
Each plot also indicates with a black line those parameter combinations $(\alpha,\beta)$ for which $\mathcal{E}_{\mathrm{LT}}=\mathcal{E}(0)$.
Except for Figure \ref{sweep}(c), these lines are barely noticeable because they coincide with $\alpha=0$. 

We remind that the average opinion can only \emph{increase}, since $d{\avg{\ln x(t)}}/{dt}>0$, Eqn.~\eqref{eq:dlnestimates}.
This was described as a rightward motion of $\avg{\ln x(t)}$ in Figure~\ref{cestationary23}. 
So, can we reach a situation where $\mathcal{E}_{\mathrm{LT}}=0$?
This depends solely on the initial condition, $\avg{\ln x(0)}$. 
Figure \ref{sweep}(a,c) depict initial configurations where $\avg{\ln x(0)} < \ln (\mathcal{T})$.
Thus, according to the plot in Figure~\ref{cestationary23}, there \emph{is} a chance that $\mathcal{E}_{\mathrm{LT}}=0$, i.e. a convergence of the collective opinions to the true value.
This is not the case for Figure \ref{sweep}(b), where the initial configuration is $\avg{\ln x(0)} > \ln (\mathcal{T})$.

Nevertheless, in all situations there are parameter sets $(\alpha,\beta)$ that minimize the collective error, $\mathcal{E}_{\mathrm{LT}}$.
Thus, we are interested to know whether this minimum value \emph{increases} or \emph{decreases} if we vary $\alpha$ or $\beta$. 
Figure \ref{sweep}(a) illustrates a situation where we have initially a \emph{large} collective error, $\mathcal{E}(0)=0.8$.
Because of this, an increase in social influence, $\alpha$, always improves the collective error.
But an increase in individual conviction $\beta$ makes this worse, because it reinforces the initial situation, which was bad. 
For $\alpha=0.4$, for instance, we see with increasing $\beta$ a color change from (green) to (red).

The situation is different in Figure \ref{sweep}(c), which as a similar initial configuration,
but a very \emph{small} initial collective error, $\mathcal{E}(0)=0.01$.
Because of this, a \emph{large} increase in social influence, $\alpha>0.5$, will lead to a deterioration.
This can be counterbalanced by an increase of individual conviction $\beta$.
Taken e.g. at $\alpha=0.8$, an increase of $\beta$ leads to a considerable improvement with respect to the collective error.

At difference with the two other configurations (a) and (b), in Figure \ref{sweep}(c) we see a \emph{non-monotonous} dependency of the collective error on the parameters $(\alpha,\beta)$.
Even that the initial collective error was very small, there are large parameter ranges where $\mathcal{E}_{\mathrm{LT}}<\mathcal{E}(0)$ (deep blue).
The collective error can also reach the minimum $\mathcal{E}_{\mathrm{LT}}=0$. 
Obviously, those parameter ranges \emph{optimize} the wisdom of crowds.
We note that that this is the case for a \emph{non-zero}, but not too large social influence, $\alpha<0.5$.

Figure \ref{sweep}(b), despite the different initial condition $\mean{\ln x(0)}>\ln \mathcal{T}$, resembles more the dependency of Figure \ref{sweep}(c) than of Figure \ref{sweep}(a), because the initial collective error is also \emph{small}, $\mathcal{E}(0)=0.02$. 
But here we do not find a non-monotonous dependency of the collective error; instead it always \emph{increases}. 
In this situation, any social influence $\alpha$ will only deteriorate the outcome, in particular if it becomes large.
This can be counterbalanced by an increasing individual conviction $\beta$. 

From this discussion we have to conclude that there is no simple monotonous impact of $\alpha$ or $\beta$ on the collective error.
It is very important how far the initial average opinion is away from the truth, and it is as important if it is below or above the true value.
Given that, there \emph{are} parameter ranges, where an \emph{increasing} social influence can also improve the collective error.
But this impact cannot be decoupled from the influence of the individual conviction, which  reinforces a good or a bad initial opinion distribution.

\begin{figure}[htbp]
  \centering
  \includegraphics[width=0.3\textwidth]{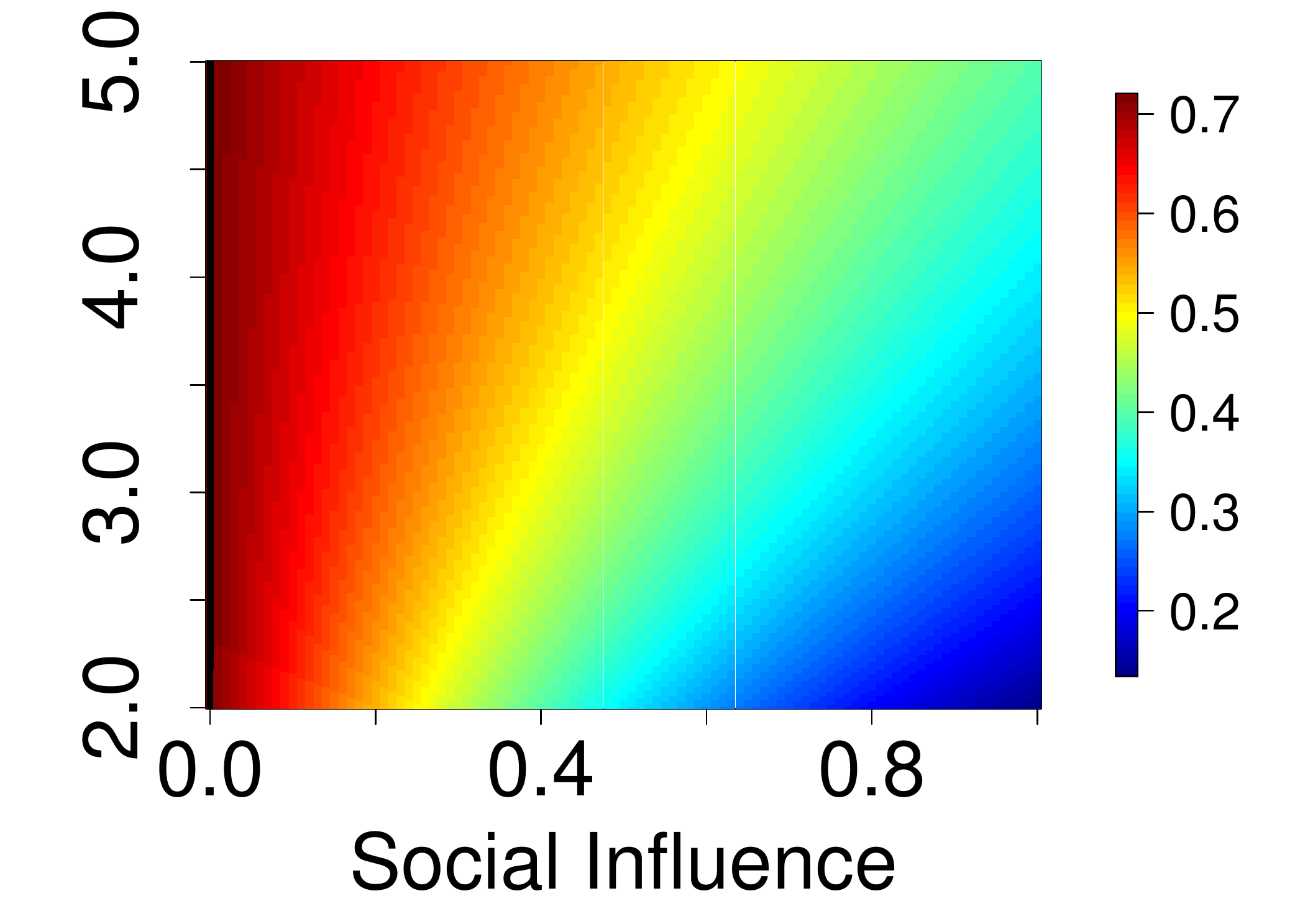}
  \caption{Color-coded long-term group diversity $D_{\mathrm{LT}}$ dependent on the values of social influence $\alpha$ ($x$-axis) and individual conviction $\beta$ ($y$-axis). 
  Initial condition: $\mathcal{D}$(0)=0.72.}
    \label{fig:diversitysweep}
\end{figure}

In Figure~\ref{fig:diversitysweep} we have plotted how the long-term group diversity depends on the parameters $(\alpha,\beta)$ for the three cases shown in Figure~\ref{sweep}.
We note that $\mathcal{D}_{\mathrm{LT}}$ behaves the same for all three cases, as it
depends only on $\alpha$, $\beta$ and $\avg{\delta^{2}(0)}$, which are
invariant across the three starting states. 
We recall that the wisdom of crowds depend on a \emph{large} group diversity, thus (red) indicates the better and (blue) the worse outcomes.
As it can be expected from the discussion above, a larger social influence $\alpha$ always negatively impacts the group diversity, regardless of the starting configuration.
Individual conviction $\beta$ acts in the opposite way - it maintains diversity in the group
by increasing the reluctance against a change of opinions.

\section{Conclusion}
\label{conclusion}

The wisdom of crowds is an intriguing phenomenon.
The observation that the \emph{average} of diverse opinions about a given questions is often more accurate than any single estimate is counter intuitive at first sight. 
But numerous anectodal, empirical and theoretical investigations across a variety of settings support this finding \cite{Surowiecki2005,Harri2008,Ray2006,Lee2011}.

However, the wisdom of crowds depends on a crucial assumption that is hard to maintain under real circumstances, namely the \emph{indepencence} of opinions.
Quite often, external events, random occurrences or social influences from others lead to a change of individual opinions that also impact the wisdom of crowds.

The aim of our paper was to study this impact of social influence in more detail.
We utilize a opinion dynamics model proposed before \citep{Mavrodiev2012a} that considers \emph{three} different ingredients: (i) the \emph{individual conviction} $\beta$ to \emph{keep} the initial opinion $x_{i}(0)$ despite other influences, (ii) the \emph{social influence} $\alpha$ to \emph{change} the own opinion $x_{i}(t)$ if information about the \emph{average opinion} $\mean{x(t)}$ becomes available, and (iii) small random influences $A\xi_{i}(t)$.

Instead of running agent-based simulations, we aimed at deriving \emph{analytic expressions} for the two most important systemic quantities that describe the wisdom of crowds effect: the collective error $\mathcal{E}(t)$ and the group diversity $\mathcal{D}(t)$.
Our only input, in addition to the opinion dynamics model, Eqn. \eqref{estimates}, are assumptions about the \emph{initial opinion distribution}, which allows us to derive expressions for the average initial opinion $\mean{\ln x(0)}$ and the average initial variance $\mean{\delta^{2}(0)}$.
A quite cumbersome derivation allowed us to find closed-form expressions for the long-term collective error $\mathcal{E}_{\mathrm{LT}}$, Eqn.~\eqref{eq:36}, and the long-term group diversity $\mathcal{D}_{\mathrm{LT}}$, Eqn.~\eqref{eq:diversity-long-term}.
These expressions could then be calculated numerically to analyze the impact of the two parameters $(\alpha,\beta)$ and the initial conditions, $\mathcal{E}(0)$, $\mathcal{D}(0)$, which have been derived from  $\mean{\ln x(0)}$ and $\mean{\delta^{2}(0)}$.

Our systematic evaluation of the impact of the parameters and the initial conditions reveals the ambiguous role of social influence $\alpha$ on the wisdom of crowds.
We could identify scenarios where increasing the social influence indeed improves the wisdom of crowds:
if the initial collective error is high, but the initial average opinion is below the true value.
But we could also demonstrate that an increasing social influence deteriorates the wisdom of crowds, if the initial collective error is already low.

This behavior is mitigated by the influence of the individual conviction $\beta$, which always reinforces the impact of the initial opinion.
In those cases, where the initial collective error is large, i.e. the starting configuration of opinions is rather bad, a large individual conviction does not improve the outcome.
But in those cases, where the initial collective error is already small, a large individual conviction helps to counterbalance the impact of social influence, and thus leads to better outcomes.

In particular, we could identify scenarios where the collective error vanishes, i.e. the average opinion converges to the true value.
This is the case if the initial collective error is low, the initial average opinion is below the true value and the social influence is at low to moderate values.

The generality of our results of course depend on the assumptions about the opinion dynamics.
It may seem that a coupling of the  individual opinion to the average opinion is not the most realistic scenario.
But this beguiles.
Analyzing experimental data, it was shown that this dynamics indeed captures the opinion dynamics of subjects in these experiments \citep{mavrodiev2013}. 
So, there is evidence for our proposed dynamics.

One could wish to generalize the derivations provided in this paper to more complex opinion dynamics that do not assume couplings to the mean.
However, we were unable to find analytic expressions for the relevant measures, collective error and group diversity, for more complex cases.
Nevertheless, the analysis provided in this paper allows us to understand and to quantify the impact of social influence on the wisdom of crowds, without the need for agent-based computer simulations.
As we have demonstrated, social influence is not inherently ``good'' or ``bad'', per se.
It depends particularly on the initial opinion distribution, i.e. the prior knowledge of the individuals, whether social influence can improve the wisdom of crowds.

\small \setlength{\bibsep}{1pt}

\appendix

\small 
\section*{Appendix}

\section{Derivation of   \boldmath{$\avg{\delta(0)\delta(t)}$}}
\label{deltasigmaappendix}

Before we can calculate our quantities of interest, we have to derive expressions for some terms involving $\delta(0)$ and $\delta(t)$ to be used later.
We remind that these $\delta$, according to Eqn.~\eqref{eq:9}, describe deviations from the mean at different times.
$\delta(0)$ is the initial deviation and follows the log-normal distribution.
$\delta(t)$, however, depends on the dynamics of $x(t)$, Eqn.~\eqref{estimates}. 

Keeping in mind that the ensemble average and the differential are linear operators and thus interchangeable:
\begin{align}
  \label{eq:12}
  \D{\avg{\delta(0)\delta(t)}}{t} &= \avg{\D{\delta(0)\delta(t)}{t}} =
  \avg{\D{\delta(t)}{t}\delta(0) + \delta(t)\D{\delta(0)}{t}}
\end{align}
The terms $d{\delta_{i}(0)}/{dt} = 0$ since all $\delta_{i}(0)$ are constants.
Hence:
\begin{align}
  \D{\avg{\delta(0)\delta(t)}}{t} &= \avg{\D{\delta(t)}{t}\delta(0)} =
  \avg{\D{(x(t)-\avg{x(t)})}{t}\delta(0)} = \avg{\D{x(t)}{t}\delta(0)}
  - \avg{\D{\avg{x(t)}}{t} \delta(0)}= \nonumber \\
  &= \avg{\D{x(t)}{t}\delta(0)} -
  \D{\avg{x(t)}}{t}\underbrace{\avg{\delta(0)}}_{=0} =
    \avg{\D{x(t)}{t}\delta(0)}
    \label{eq:13}
\end{align}
Now we can plug in $x(t)$ from Eqn.~\eqref{estimates}:
\begin{align}
  \label{deltasigmaappendix0}
  \D{\avg{\delta(0)\delta(t)}}{t} =&
  \avg{\delta(0)\alpha(\underbrace{\avg{x(t)}-x(t)}_{=-\delta(t)})
    + \delta(0)\beta(x(0)-x(t)) +
    A\xi(t)\delta(0)}  \nonumber \\
  =& -\alpha\avg{\delta(t)\delta(0)} +
  \beta\avg{\delta(0)(\avg{x(0)}+\delta(0))} -
  \beta\avg{\delta(0)(\avg{x(t)}+\delta(t))} + \nonumber  \\
  &+\dfrac{A}{\sqrt{N}}\avg{\xi(t)\delta(0)} \nonumber  \\
  =& -\alpha\avg{\delta(t)\delta(0)} +
  \beta\avg{\delta^{2}(0)}-\beta\avg{\delta(0)\delta(t)} + \dfrac{A}{\sqrt{N}}\underbrace{\avg{\xi(t)\delta(0)}}_{=\avg{\xi(t)}\avg{\delta(0)}=0} \nonumber  \\
  =& \beta\avg{\delta^{2}(0)}-(\alpha+\beta)\avg{\delta(0)\delta(t)}
\end{align}
where we have used the fact that $\xi_{i}(t)$ and $\delta_{i}(0)$ have
negligible covariance. This is true because generation of the
$\delta_{i}$ does not depend on the white noise, nor does the
$\xi_{i}$'s depend on the initial distribution of $\delta_{i}$. The
solution of the above equation is:
\begin{align}
  \label{eq:14}
  \avg{\delta(0)\delta(t)} = \dfrac{\beta\avg{\delta^{2}(0)}}{\alpha+\beta} + \text{C}e^{-(\alpha+\beta)t}
\end{align}
The constant C is given by the initial condition
$\avg{\delta(0)\delta(0)}$:
\begin{align}
  \label{eq:15}
  \text{C} = \avg{\delta(0)\delta(0)} -
  \dfrac{\beta\avg{\delta^{2}(0)}}{\alpha+\beta}
\end{align}
However, $\avg{\delta(0)\delta(0)} = \avg{\delta^{2}(0)}$, hence:
\begin{align}
  \label{constantC}
  \text{C} = \dfrac{\alpha}{\alpha+\beta}\avg{\delta^{2}(0)}
\end{align}
Therefore:
\begin{align}
  \avg{\delta(0)\delta(t)} &=
  \dfrac{\beta\avg{\delta^{2}(0)}}{\alpha+\beta}
  +\dfrac{\alpha}{\alpha+\beta}\avg{\delta^{2}(0)}e^{-(\alpha+\beta)t}
\label{deltadelta2}
\end{align}

\section{Derivation of \boldmath{$\avg{\delta(0)\delta^{2}(t)}$}}

We use again the fact that the ensemble average and the differential are interchangeable:
\begin{align}
  \label{eq:16}
  \D{\avg{\delta(0)\delta^{2}(t)}}{t} &= \avg{\D{\delta(0)\delta^{2}(t)}{t}} =
  \avg{\delta(0)\D{\delta^{2}(t)}{t} + \delta^{2}(t)\D{\delta(0)}{t}}
\end{align}
The $\delta(0)$ are constant, hence $d{\delta(0)}/{dt} = 0$:
\begin{align}
  \D{\avg{\delta(0)\delta^{2}(t)}}{t} &=
  2\avg{\delta(0)\delta(t)\D{\delta(t)}{t}}=2\avg{\delta(0)\delta(t)\D{}{t}\left(x(t)-\avg{x(t)}\right)} \nonumber \\
  &=2\Bigg[
  \avg{\delta(0)\delta(t)\D{}{t}x(t)} - \avg{\delta(0)\delta(t)}\D{}{t}\avg{x(t)}\Bigg]
\end{align}
Plugging in $x(t)$ from  Eq.~\ref{estimates} yields:
\begin{align}
  \D{\avg{\delta(0)\delta^{2}(t)}}{t} &= 2\Bigg[
  \avg{\delta(0)\delta(t)\Big[ \alpha
    (\underbrace{\avg{x(t)}-x(t)}_{{\scriptscriptstyle
        =-\delta(t)}})+\beta(x(0)-x(t))+A\xi(t) \Big]}
  -\avg{\delta(0)\delta(t)}\D{}{t}\avg{x(t)}\Bigg] \nonumber \\
&=-2\alpha\avg{\delta(0)\delta^{2}(t)}+\hcancel{2
  \beta\avg{\delta(0)\delta(t)}\big( \avg{x(0)}-\avg{x(t)} \big)} + 2\beta\avg{\delta^{2}(0)\delta(t)} \\
&\quad -2\beta\avg{\delta(0)\delta^{2}(t)}+\dfrac{2A}{\sqrt{N}}\avg{\xi(t)}
-\hcancel{2\beta\avg{\delta(0)\delta(t)}\big(\avg{x(0)}-\avg{x(t)}}-\dfrac{2A}{\sqrt{N}}\avg{\xi(t)}
\big) \nonumber
\end{align}
The noise term becomes negligible for large $N$, hence we can rewrite:
\begin{equation}
\label{eq:53}
\D{\avg{\delta(0)\delta^{2}(t)}}{t} = 2\beta\avg{\delta^{2}(0)\delta(t)}-2(\alpha+\beta)\avg{\delta^{2}(0)\delta(t)}
\end{equation}
Making use of the expression for $\avg{\delta^{2}(0)\delta(t)}$, Eqn.~\eqref{eq:18}, derived below, we can rewrite  
Eqn.~\eqref{eq:53} as:
\begin{align}
  \D{\avg{\delta(0)\delta^{2}(t)}}{t} &=- 2(\alpha+\beta)\avg{\delta^{2}(0)\delta(t)}+\dfrac{2\beta^{2}}{\alpha+\beta}\avg{\delta^{3}(0)}+\dfrac{2\alpha\beta}{\alpha+\beta}\avg{\delta^{3}(0)}e^{-(\alpha+\beta)t}
                                        \label{eq:20}
\end{align}
which has the closed-form solution:
\begin{equation}
\avg{\delta(0)\delta^{2}(t)}
=\dfrac{\avg{\delta^{3}(0)}}{(\alpha+\beta)^{2}}\Big[\beta^{2}+2\alpha\beta
e^{-(\alpha+\beta)t}+\alpha^{2}e^{-2(\alpha+\beta)t}\Big]
\label{eq:19}
\end{equation}
This can be generalized as follows:
\begin{equation}
  \label{eq:s0dtn}
  \avg{\delta(0)\delta^{n}(t)} =
  \dfrac{\avg{\delta^{n+1}(0)}}{(\alpha+\beta)^{n}}\big[\beta+\alpha e^{-(\alpha+\beta)t}\big]^{n}
\end{equation}

\section{Derivation of \boldmath{$\avg{\delta^{2}(0)\delta(t)}$}}
\label{sec:deriv-boldm}

We can use the same method applied in the previous section to expand:
\begin{align}
  \label{eq:21}
  \D{\avg{\delta^{2}(0)\delta(t)}}{t} &=
  \avg{\delta(t)\underbrace{\D{}{t}\delta^{2}(0)}_{{\scriptscriptstyle =
        0}} + \delta^{2}(0)\D{}{t}\delta(t)} =
  \avg{\delta^{2}(0)\D{}{t}\big(x(t)-\avg{x(t)}\big)} \nonumber \\
&= \avg{\delta^{2}(0)\D{}{t}x(t)} - \avg{\delta^{2}(0)}\D{}{t}\avg{x(t)}
\end{align}
Using the expressions for $x(t)$ and $\avg{x(t)}$ from Eqs.~\eqref{estimates}, \eqref{estimates2}, we obtain:
\begin{align}
  \D{\avg{\delta^{2}(0)\delta(t)}}{t} &=
  \avg{\delta^{2}(0)\Big[\alpha\big(\avg{x(t)}-x(t)\big) +
    \beta\big(x(0)-x(t)\big) + A\xi(t)\Big]} - \nonumber \\
&\quad -
\avg{\delta^{2}(0)}\Big[\beta\big(\avg{x(0)}-\avg{x(t)}\big)+\dfrac{A}{\sqrt{N}}\avg{\xi(t)}\Big]
 \nonumber \\
&=
-\alpha\avg{\delta^{2}(0)\delta(t)}+\hcancel{\beta\avg{\delta^{2}(0)}\big(\avg{x(0)}-\avg{x(t)}\big)}+\beta\avg{\delta^{3}(0)}-\beta\avg{\delta^{2}(0)\delta(t)}+\nonumber \\
&\quad +
D\avg{\delta^{2}(0)\xi(t)}-\hcancel{\beta\avg{\delta^{2}(0)}\big(\avg{x(0)}-\avg{x(t)}\big)}-\dfrac{A}{\sqrt{N}}\avg{\delta^{2}(0)}\avg{\xi(t)}
 \nonumber \\
&= \beta\avg{\delta^{3}(0)}-(\alpha+\beta)\avg{\delta^{2}(0)\delta(t)}
       \label{eq:17}
\end{align}
where we have used that the expectations involving the noise terms are
0. The solution to this first-order ODE is given by:
\begin{align}
  \label{eq:18}
  \avg{\delta^{2}(0)\delta(t)}
  =\dfrac{\beta}{\alpha+\beta} \avg{\delta^{3}(0)}+\dfrac{\alpha}{\alpha+\beta}\avg{\delta^{3}(0)}e^{-(\alpha+\beta)t}
\end{align}
Using the same line of arguments, we can obtain: 
\begin{equation}
\label{eq:s0mdtn}
\avg{\delta^{m}(0)\delta^{n}(t)} =
\dfrac{\avg{\delta^{m+n}(0)}}{(\alpha+\beta)^{n}}\big[\beta+\alpha
e^{-(\alpha+\beta)t}\big]^{n}, \quad\quad m,n \ge 0
\end{equation}
which is a generalization of the previous Eqns.~\eqref{deltadelta2}, \eqref{eq:s0dtn}, \eqref{eq:s0mdtn}.

\section{Derivation of \boldmath{$\avg{\delta^{2}(t)}$}}
\label{deltasqappendix}

Eventually, we express: 
\begin{align}
  \label{eq:22}
  \D{\avg{\delta^{2}(t)}}{t} &= 2\avg{\delta(t)\D{\delta(t)}{t}}
\end{align}
Rewriting $\delta(t)$ from Eqn.~\eqref{eq:9}, and using Eqs.~\eqref{estimates}, 
\eqref{estimates2} for $x(t)$ and $\avg{x(t)}$, respectively:
\begin{align}
  &2\avg{\delta(t)\D{\delta(t)}{t}} =
  2\avg{\delta(t)\left(\D{x(t)}{t}-\D{\avg{x(t)}}{t}\right)} 
  \nonumber \\
  &=2\avg{\delta(t)\left(\alpha(\underbrace{\avg{x(t)}-x(t)}_{=-\delta(t)})+\beta\big[\underbrace{x(0)
    -\avg{x(0)}}_{=\delta(0)}+\underbrace{\avg{x(t)}-x(t)}_{=-\delta(t)}\big]
    +D\left(\xi(t)-\dfrac{1}{\sqrt{N}}\avg{\xi(t)}\right)\right)}
   \nonumber\\
  &=2\avg{\delta(t)\left(-\alpha\delta(t)+\beta(\delta(0)-\delta(t))+D\left(\xi(t)-\dfrac{1}{\sqrt{N}}\avg{\xi(t)}\right)\right)}
  \nonumber\\
&=2\beta\avg{\delta(0)\delta(t)}-2(\alpha+\beta)\avg{\delta^{2}(t)}+2\dfrac{A}{\sqrt{N}}\avg{\delta(t)\xi(t)}-2\dfrac{A}{\sqrt{N}}\avg{\xi(t)}\avg{\delta(t)}
                \label{eq:23}
\end{align}
$\xi_{i}(t)$ is Gaussian white noise, and we assume that it has a
negligible influence on $x_{i}(t)$. This is a reasonable assumption,
because the major contribution to $d{x_{i}(t)}/{dt}$ comes from $\alpha$
and $\beta$, hence it is justified to conclude that $x_{i}(t)$ and
$\xi_{i}(t)$ have a negligible covariance. Hence
\mbox{$\avg{\delta(t)\xi(t)}-\avg{\delta(t)\avg{\xi(t)}} \approx 0$}. Using this
we obtain:
\begin{align}
  \label{ddeltatsq}
  \D{\avg{\delta^{2}(t)}}{t} &=
  2\beta\avg{\delta(0)\delta(t)}-2(\alpha+\beta)\avg{\delta^{2}(t)}
\end{align}
Plugging in Eqn.~\eqref{deltadelta2}, the closed form solution is given by:
\begin{align}
  \avg{\delta^{2}(t)} =
  \dfrac{\beta^{2}\avg{\delta^{2}(0)}}{(\alpha+\beta)^{2}} +
  \dfrac{2\text{C}\beta e^{-(\alpha+\beta)t}}{\alpha+\beta} +
  \text{C}_{1}e^{-2(\alpha+\beta)t}
\end{align}
where $C_{1}$ is a constant obtained from
$\avg{\delta^{2}(t=0)}=\avg{\delta^{2}(0)}$, and $C$ is
the constant from \ref{constantC}. Substituting the constants yields:
\begin{align}
  \avg{\delta^{2}(t)} &=\dfrac{\left[e^{-(\alpha +\beta)t} \alpha +\beta
    \right]^2 \avg{\delta^{2}(0)} }{(\alpha +\beta )^2}
                        \label{eq:24}
\end{align}
Again, this can be generalized to: 
\begin{equation}
  \label{eq:dtn}
  \avg{\delta^{n}(t)} =
  \dfrac{\avg{\delta^{n}(0)}}{(\alpha+\beta)^{n}}\big[\beta+\alpha e^{-(\alpha+\beta)t}\big]^{n}
\end{equation}

\section{Derivation of Eqn.~\eqref{eq:dlnestimates}}
  \label{dlogestimates2appendix}

We start from Eqn.~\eqref{estimates}, which can be rewritten as:
\begin{align}
  \label{dlogestimate}
  \D{\ln x_{i}(t)}{t}
  &= \dfrac{\alpha \avg{x(t)}}{x_{i}(t)} -\alpha + \dfrac{\beta
      x_{i}(0)}{x_{i}(t)} - \beta + \dfrac{A\xi_{i}(t)}{x_{i}(t)}
\end{align}
The ensemble average of Eqn.~\eqref{dlogestimate} is:
\begin{align}
  \label{dlogestimates}
  \D{\avg{\ln x(t)}}{t} &= \alpha\avg{\dfrac{\avg{x(t)}}{x_{i}(t)}} -
                          \alpha + \beta\avg{\dfrac{x_{i}(0)}{x_{i}(t)}} - \beta +
                          \dfrac{A}{\sqrt{N}}\avg{\dfrac{\xi_{i}(t)}{x_{i}(t)}}
\end{align}
$x_{i}(t)$ and $x_{i}(0)$ can be expressed by small deviations from their respective means, as written in Eqn.~\eqref{eq:9}. 
We further use a Taylor expansion around zero:
\begin{align}
  \label{eq:25}
  \left(1+\dfrac{\delta(t)}{\avg{x(t)}}\right)^{-1}= \sum_{n=0}^{\infty} \dfrac{(-1)^{n}}{\avg{x(t)}^{n}}\avg{\delta^{n}(t)}
\end{align}
With this, Eqn.\ref{dlogestimates} can be approximated as follows:
\begin{align}
    \D{\avg{\ln x(t)}}{t} &=
                            \alpha\avg{
                            {\left(1+\dfrac{\delta(t)}{\avg{x(t)}}\right)^{-1}}}
                            - \alpha + \beta\avg{\dfrac{x(0)}{x(t)}}
    - \beta +
    \dfrac{A}{\sqrt{N}}\avg{\dfrac{\xi(t)}{x(t)}} \nonumber \\
    &=\alpha\avg{\sum_{n=0}^{\infty} \left(-\dfrac{\delta(t)}{\avg{x(t)}}\right)^{n}}+\beta\avg{\dfrac{\avg{x(0)}\left(1+\dfrac{\delta(0)}{\avg{x(0)}}\right)}{\avg{x(t)}\left(1+\dfrac{\delta(t)}{\avg{x(t)}}\right)}}
    - \alpha -\beta
    +\dfrac{A}{\sqrt{N}}\avg{\dfrac{\xi(t)}{x(t)}} \nonumber  \\
    &=\alpha\avg{\sum_{n=0}^{\infty} \dfrac{(-1)^{n}}{\avg{x(t)}^{n}}\avg{\delta^{n}(t)}}+
      \beta\dfrac{\avg{x(0)}}{\avg{x(t)}}\avg{\left(1+\dfrac{\delta(0)}{\avg{x(0)}}\right)
\left(1+\dfrac{\delta(t)}{\avg{x(t)}}\right)^{-1}}
      \nonumber    \\
    &\quad -\alpha-\beta+\dfrac{A}{\sqrt{N}}\avg{\dfrac{\xi(t)}{x(t)}} =\nonumber 
    \\
    &=\alpha\sum_{n=0}^{\infty} \dfrac{(-1)^{n}}{\avg{x(t)}^{n}}\avg{\delta^{n}(t)}+\beta\dfrac{\avg{x(0)}}{\avg{x(t)}}\avg{\left(1+\dfrac{\delta(0)}{\avg{x(0)}}\right).\sum_{n=0}^{\infty} \dfrac{(-1)^{n}}{\avg{x(t)}^{n}}\avg{\delta^{n}(t)}}
    -  \nonumber  \\
    &\quad -\alpha-\beta+\dfrac{A}{\sqrt{N}}\avg{\dfrac{\xi(t)}{x(t)}}
    \nonumber \\
    &=\alpha\sum_{n=0}^{\infty}
      \dfrac{(-1)^{n}}{\avg{x(t)}^{n}}\avg{\delta^{n}(t)} + \beta\dfrac{\avg{x(0)}}{\avg{x(t)}}\left[\sum_{n=0}^{\infty}
      \dfrac{(-1)^{n}}{\avg{x(t)}^{n}}\avg{\delta^{n}(t)}
    + \sum_{n=0}^{\infty}
      \dfrac{(-1)^{n}\avg{\delta(0)\delta^{n}(t)}}{\avg{x(0)}\avg{x(t)}^{n}}\right]- \nonumber  \\
    &\quad -\alpha-\beta+\dfrac{A}{\sqrt{N}}\avg{\dfrac{\xi(t)}{x(t)}}
  \end{align}
We now express $\avg{\delta^{n}(t)}$ by means of Eqn.~\eqref{eq:dtn} and
$\avg{\delta(0)\delta^{n}(t)}$ by means of Eqn.~\eqref{eq:s0mdtn}. 
Further, we make again the assumption that the random noise, $\xi_{i}(t)$ is
negligibly small and not correlated to $x_{i}(t)$.
As a consequence:
\begin{align}
  \label{eq:26}
  \avg{\dfrac{\xi(t)}{x(t)}} = \dfrac{\avg{\xi(t)}}{\avg{x(t)}}
\end{align}
Hence, we obtain:
\begin{align}
    \D{\avg{\ln x(t)}}{t} & =\big(\alpha+\beta)\sum_{n=0}^{\infty}
      \dfrac{(-1)^{n}}{\avg{x(0)}^{n}} \dfrac{\avg{\delta^{n}(0)}}{(\alpha+\beta)^{n}}\big[\beta+\alpha
      e^{-(\alpha+\beta)t}\big]^{n} + \nonumber \\
&\quad +\beta \sum_{n=0}^{\infty}
      \dfrac{(-1)^{n}}{\avg{x(0)}^{n+1}} \dfrac{\avg{\delta(0)^{n+1}}}{(\alpha+\beta)^{n}}\big[\beta+\alpha
                                                     e^{-(\alpha+\beta)t}\big]^{n}
                                                     - (\alpha+\beta) +\dfrac{A}{\sqrt{N}}\dfrac{\avg{\xi(t)}}{\avg{x(t)}} 
\end{align}
where we have used that $\avg{x(t)} = \avg{x(0)}$ for large $t$.
We can transform the second term in the following way:
\begin{align}
  \label{eq:27}
& \beta \sum_{n=0}^{\infty}
      \dfrac{(-1)^{n}}{\avg{x(0)}^{n+1}} \dfrac{\avg{\delta(0)^{n+1}}}{(\alpha+\beta)^{n}}\big[\beta+\alpha
                                                     e^{-(\alpha+\beta)t}\big]^{n} \nonumber \\
&=  \dfrac{\beta (\alpha+\beta)}{\beta+\alpha e^{-(\alpha+\beta)t}} \sum_{n=0}^{\infty}
      \dfrac{(-1)^{n+1}}{\avg{x(0)}^{n+1}} \dfrac{\avg{\delta(0)^{n+1}}}{(\alpha+\beta)^{n+1}}\big[\beta+\alpha
       e^{-(\alpha+\beta)t}\big]^{n+1} \nonumber \\
& =\dfrac{\beta (\alpha+\beta)}{\beta+\alpha e^{-(\alpha+\beta)t}} \sum_{n=1}^{\infty}
       \dfrac{(-1)^{n}}{\avg{x(0)}^{n}} \dfrac{\avg{\delta^{n}(0)}}{(\alpha+\beta)^{n}}\big[\beta+\alpha
       e^{-(\alpha+\beta)t}\big]^{n}
\end{align}
This leads to the dynamics finally written in Eqn.~\eqref{eq:dlnestimates}.

\section{Derivation of Eqn.~\eqref{eq:3}}
\label{sec:derivation-eqn}
Finally, we have to integrate the dynamics of Eqn.~\eqref{eq:dlnestimates}: 
\begin{align}
\dfrac{1}{\alpha+\beta}\int \! \D{\avg{\ln x(t)}}{t} \mathrm{d}t &= 
\sum_{n=1}^{\infty}
      \dfrac{(-1)^{n}}{\avg{x(0)}^{n}} \dfrac{\avg{\delta^{n}(0)}}{(\alpha+\beta)^{n}}\underbrace{\int \!\big[\beta+\alpha
      e^{-(\alpha+\beta)t}\big]^{n} \mathrm{d}t}_{:=~\mathcal{A}(n,t)} - \nonumber \\
&\quad - \beta 
\sum_{n=1}^{\infty}
      \dfrac{(-1)^{n}}{\avg{x(0)}^{n}} \dfrac{\avg{\delta^{n}(0)}}{(\alpha+\beta)^{n}}\int \!\big[\beta+\alpha
      e^{-(\alpha+\beta)t}\big]^{n-1} \mathrm{d}t + \nonumber \\
&\quad + \dfrac{A}{\avg{x(0)}\sqrt{N}} \displaystyle \int_{0}^{t}e^{\beta(s-t)}\avg{\xi(s)}ds
\label{eq:28}
\end{align}
The integral $\mathcal{A}(n,t)$ yields:
\begin{align}
  \mathcal{A}(n,t) &= \beta^{n}t + C + \sum_{k=1}^{n}{n \choose
    k}\beta^{n-k}\alpha^{k} \int \! e^{-(\alpha+\beta)tk}\mathrm{d}t =
  \nonumber \\
&=\beta^{n}t + C - \sum_{k=1}^{n}{n \choose
    k}\beta^{n-k}\alpha^{k}\left[\dfrac{1}{(\alpha+\beta)k}e^{-(\alpha+\beta)kt}
    - C\right] = \nonumber \\
&=\beta^{n}t + C\left[1+(\alpha+\beta)^{n}-\beta^{n}\right]-\dfrac{1}{\alpha+\beta}\sum_{k=1}^{n}\dfrac{1}{k}{n \choose
    k}\beta^{n-k}\alpha^{k}e^{-(\alpha+\beta)kt}
\end{align}
Now we can write $\mathcal{B}(n,t):=\mathcal{A}(n,t)-\beta \mathcal{A}(n-1,t)$ and express the solution of Eqn.~\eqref{eq:28} as:
\begin{align}
 \avg{\ln x(t)} &= \sum_{n=1}^{\infty}
      \dfrac{(-1)^{n}}{\avg{x(0)}^{n}}
      \dfrac{\avg{\delta^{n}(0)}}{(\alpha+\beta)^{n-1}} \mathcal{B}(n,t)+\dfrac{A(\alpha+\beta)}{\avg{x(0)}\sqrt{N}} \displaystyle \int_{0}^{t}e^{\beta(s-t)}\avg{\xi(s)}ds
                  \label{eq:29}
\end{align}
$\mathcal{B}(n,t)$ equals:
\begin{align}
\mathcal{B}(n,t) &=\hcancel{\beta^{n}t} + C\left[1+(\alpha+\beta)^{n}-\beta^{n}\right]-\dfrac{1}{\alpha+\beta}\sum_{k=1}^{n}\dfrac{1}{k}{n \choose
    k}\beta^{n-k}\alpha^{k}e^{-(\alpha+\beta)kt}-\nonumber \\
&\quad - \hcancel{\beta^{n}t} - \beta C \left[1+(\alpha+\beta)^{n-1}-\beta^{n-1}\right]+\dfrac{\beta}{\alpha+\beta}\sum_{k=1}^{n-1}\dfrac{1}{k}{n-1 \choose
    k}\beta^{n-1-k}\alpha^{k}e^{-(\alpha+\beta)kt}
\label{eq:30}
\end{align}
Using the property of the binomial coefficient that ${n \choose k} = {n
  \choose k-1} + {n-1 \choose k-1}$ we write further:
\begin{align}
\mathcal{B}(n,t) &= C\big[1-\beta+\alpha(\alpha+\beta)^{n-1}\big] -
\dfrac{\alpha^{n}}{n(\alpha+\beta)}e^{-(\alpha+\beta)nt} - \dfrac{1}{\alpha+\beta}\sum_{k=1}^{n-1}\dfrac{1}{k}{n-1 \choose  k-1}\beta^{n-k}\alpha^{k}e^{-(\alpha+\beta)kt}
                   \label{eq:31}
\end{align}
with 
\begin{align}
  \mathcal{B}(n,\infty) &= C\big[1-\beta+\alpha(\alpha+\beta)^{n-1}\big] \nonumber \\
  \mathcal{B}(n,0) &= C\big[1-\beta+\alpha(\alpha+\beta)^{n-1}\big] - \dfrac{\alpha^{n}}{n(\alpha+\beta)}- \dfrac{1}{\alpha+\beta}\sum_{k=1}^{n-1}\dfrac{1}{k}{n-1 \choose
    k-1}\beta^{n-k}\alpha^{k}
                     \label{eq:33}
\end{align}
The first term involves the integration constant, $C$, and  can be determined from the initial condition
$\avg{x(0)}$:
\begin{align}
& \sum_{n=1}^{\infty}
      \dfrac{(-1)^{n}}{\avg{x(0)}^{n}}
      \dfrac{\avg{\delta^{n}(0)}}{(\alpha+\beta)^{n-1}}
                \Bigg(C\big[1-\beta+\alpha(\alpha+\beta)^{n-1}\big]\Bigg) \nonumber \\
&  =      \avg{\ln x(0)} + \sum_{n=1}^{\infty}
      \dfrac{(-1)^{n}}{\avg{x(0)}^{n}}
      \dfrac{\alpha^{n}\avg{\delta^{n}(0)}}{n(\alpha+\beta)^{n}} +
\sum_{n=1}^{\infty}
      \dfrac{(-1)^{n}}{\avg{x(0)}^{n}}
      \dfrac{\avg{\delta^{n}(0)}}{(\alpha+\beta)^{n}}\sum_{k=1}^{n-1}\dfrac{1}{k}{n-1 \choose
    k-1}\beta^{n-k}\alpha^{k}
                     \label{eq:34}
\end{align}
where we have also zeroed the stochastic term, which is true for large
$N$. Hence finally:
\begin{align}
  \avg{\ln x(t)} &= \avg{\ln x(0)}
  + \sum_{n=1}^{\infty}
      \dfrac{(-1)^{n}}{\avg{x(0)}^{n}}
                    \dfrac{\alpha^{n}\avg{\delta^{n}(0)}}{n(\alpha+\beta)^{n}}
                     +       \nonumber \\
&\quad +\sum_{n=1}^{\infty}
      \dfrac{(-1)^{n}}{\avg{x(0)}^{n}}
      \dfrac{\avg{\delta(0)^{n}}}{(\alpha+\beta)^{n}}\sum_{k=1}^{n-1}\dfrac{1}{k}{n-1 \choose
                                            k-1}\beta^{n-k}\alpha^{k} -\sum_{n=1}^{\infty}
      \dfrac{(-1)^{n}}{\avg{x(0)}^{n}}
      \dfrac{\alpha^{n}\avg{\delta^{n}(0)}}{n(\alpha+\beta)^{n}}e^{-(\alpha+\beta)nt}
      \nonumber \\
&\quad -\sum_{n=1}^{\infty}
      \dfrac{(-1)^{n}}{\avg{x(0)}^{n}}
      \dfrac{\avg{\delta^{n}(0)}}{(\alpha+\beta)^{n}}\sum_{k=1}^{n-1}\dfrac{1}{k}{n-1 \choose
    k-1}\beta^{n-k}\alpha^{k}e^{-(\alpha+\beta)kt}
                  \label{eq:35}
\end{align}
or more compactly as written in Eqn.~\eqref{eq:3}.

\end{document}